\documentclass[fleqn,twoside]{article}
\usepackage{latexsym}
\usepackage{espcrc2}
\readRCS
$Id: espcrc2.tex,v 1.2 2004/02/24 11:22:11 spepping Exp $
\usepackage{graphicx}

\newcommand{\AmS}{{\protect\the\textfont2
  A\kern-.1667em\lower.5ex\hbox{M}\kern-.125emS}}

\title{
\vspace{-5mm}
\rightline{\small DESY 05-155,  RCNP-Th05028}
Optimization of  Lattice QCD codes for the AMD Opteron processor}

\author{Miho Koma\address[DESY]{DESY Theory Group, 
        Notkestrasse 85, D-22603 Hamburg, Germany}
	\address{RCNP Osaka University, Mihogaoka 10-1, 
	Ibaraki, Osaka 567-0047, Japan}
}                                                  

\begin{document}

\begin{abstract}
    We report our experience of the optimization of the lattice QCD 
    codes for the new Opteron cluster at DESY Hamburg,
    including benchmarks.
    Details of the optimization using SSE/SSE2 instructions and 
    the effective use of prefetch instructions are discussed.
\vspace{1pc}
\end{abstract}

\maketitle

\section{Introduction}
Lattice QCD is a powerful method to study Quantum chromodynamics 
 (QCD) in a nonperturbative way.
In lattice QCD, a path integral
is directly evaluated
on a discrete space-time lattice 
by means of 
the Monte Carlo method.
As computer technology advances 
PC clusters can also be used for lattice QCD simulations
as well as a number of commercial supercomputers.

Since  lattice QCD simulations demand huge computer power,
it is very important to optimize the simulation codes so as to 
exploit the full potential of the processor.
Thus we optimize the hot spots of the codes such as  
the operation of a
Dirac operator to a spinor (referred as $Q\phi$ hereafter) and
linear algebra of 
spinors, e.g. 
\begin{itemize}
    \setlength{\itemsep}{0pt}
    \setlength{\parsep}{0pt}
    \item  $\langle\psi,\phi\rangle$: scalar product of two spinors
    \item  $\langle\psi,\psi\rangle$: norm square of a spinor 
    \item  $\psi=\psi+c\phi$: add two spinors with a constant factor and assign the result to
one of the source spinors
\end{itemize}
In the simulation, a spinor is defined as a vector with 12 complex 
components on each grid point (site) of the lattice, 
and a gauge field
is defined as a complex $3\times 3$ matrix 
on a link, which connects nearest neighbor sites.
Numerically the operation $Q\phi$
is a combination of a complex $3\times 3$ matrix times a 
complex vector.
From a technical point of view,
we should optimize these two applications separately since
 $Q\phi$ is processor limited 
as it needs more than one thousand floating point
operations per site, while the linear algebra routines
demand more data transfer than actual 
computations (memory limited).

In this note we report our experience of the optimization of the
lattice QCD codes for the new PC cluster with the AMD Opteron
processor~\cite{man} which is recently installed at DESY Hamburg.  Of
course the parallel computing with multiple processors is of our
interest, however, we have optimized here only single processor
version as a first step. 
We also compare benchmarks with an older PC
cluster with the Intel Xeon processors 
(See Table~\ref{tbl:spec} for
the specification of these two processors).

\begin{table}
    \caption{Comparison of the specification of the two processors 
    which we use}
    \small
    \begin{tabular}{ccc}
	\hline
	& AMD Opteron & Intel Xeon  \\
	\hline
	Processor speed & 2.4 GHz & 1.7 GHz  \\
	L1 Data Cache & 64 kbytes & ---  \\
	L2 Data Cache & 1024 kbytes & 512 kbytes  \\
	\# of SSE registers & 16 & 8  \\
	Cache line size & 64 bytes & 128 bytes \\
	\hline
    \end{tabular}
    \label{tbl:spec}
\end{table}

\section{SSE instructions}
To achieve the full performance, it is important to make use of
the special feature of the processor.
Streaming SIMD extensions instruction sets
(SSE) is the case for us~\cite{sse}.
SSE is designed to process 128-bit long data which may contain
multiple elements of vectors
and is suitable for vector operations like $Q\phi$ and linear algebra.
Both Opteron and Xeon processors support SSE and SSE2 (the first 
extension of SSE) and have
special 128-bit long registers for the SSE instructions (SSE registers)
as summarized in Table~\ref{tbl:spec}.
Note that neither of them supports SSE3, the latest extension of SSE.

In our group, simulation codes written in the standard C language 
with SSE/SSE2  have already been developed
for the Xeon processor.
These SSE instructions are embedded to the C codes by defining 
macros using GCC inline assembly~\cite{inline}.
The macro contains the SSE instructions that load data 
from system memory to 
a SSE register,  operate on vectors on the SSE registers, 
store data to 
system memory, etc. 

Technically it is also important to consider the latency of the SSE
instruction itself as well as the latency caused by the data transfer.
Usually an instruction needs several processor cycles to finish the
operation.  Although the Opteron processor can execute up to three
instructions per cycle, the processor has to wait until the previous
operation is finished if these two operations are interdependent.  One
can avoid this by performing independent operations by using several
SSE registers.  For instance, in a macro with SSE instructions, we
first load $512$-bit data to four SSE registers per function call.
Then the operation on the first SSE register starts while data loading
proceeds on the third and the fourth registers.  In the same way, the
data on the first SSE register can be stored to the system memory
while the calculation on the third and the fourth registers are going
on.  This is how one can hide the latency of the SSE instructions.  In
addition to this, of course, the reduction of the number of
instructions and/or the use of lower latency instruction improves
performance.  Opteron's twice the legacy number of registers (see
Table~\ref{tbl:spec}) can eliminate substantial memory access overhead
by keeping intermediate results on SSE registers and can hide latency
of each instruction more effectively.

\section{Prefetch instruction}
The mismatch between the memory bandwidth
and the processor throughput is one of the big origins of latency.
Typical difference of them is a factor of ten. 
The prefetch instructions, 
which read data from system memory into the 
level~1 (L1) data cache,
take advantage of the high bus bandwidth of 
the Opteron processor to hide latencies when fetching the data from 
the system memory.
Data transfer is processed in background and
eight prefetch instructions can be ``in flight'' at a time.
As a prefetch instruction initiates a read request of a 
specified address and reads one entire cache line that contains the
requested address,
prefetch instructions can improve performance 
in situations where the sequential addresses are read.
This is the case for lattice QCD 
simulations.
Once the data is in the L1 cache, there is almost no latency as
128-bit data can be loaded into a SSE 
register within two processor cycles.
Note that the number of prefetch instructions
is dependent on the length of one cache line, which is 64 
bytes for the Opteron processor, 
while 128 bytes for the Xeon processor. 

One of the important parameter
for the effective use of prefetch instructions
is ``prefetch distance,'' which 
denotes how far ahead the prefetch request is made.
In principle, the prefetch distance should be long enough so that the 
data is in the cache by the time it is needed by the processor.
The actual distance is dependent on the application.

\begin{figure}[t]
    \includegraphics[width=7.9cm]{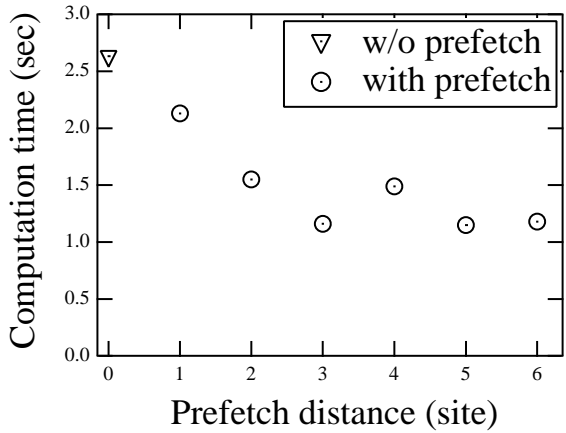}
    \caption{Prefetch distance dependence of the performance
    of the macro which computes $3\times 3$ 
    complex matrix times complex vector on each site on 
    $12^{3}\times 24$ lattice with 16 SSE registers, which is
    optimized for 
    the Opteron processor. The computation 
    time is shown in units of second for 500 applications.
    Prefetch distance ``1'' means that a spinor  on 
    the next site and a gauge field on the next link 
    is prefetched before the computation on a certain site.
    }
\label{fig:prefetch}
\end{figure}

We show the
prefetch distance dependence of the
performance of the macro which computes a $3\times 3$ 
complex matrix times a complex vector on each site in 
Fig.~\ref{fig:prefetch}.
This operation can be regarded as a prototype of $Q\phi$.
We find that the macro with the prefetch distance 
three is more than twice faster than that without prefetch instruction. 
It is also interesting to note that shorter prefetch 
distance than the optimal distance (distances 1 or 2 in 
Fig~\ref{fig:prefetch})
gives worse performance compare to 
longer prefetch distance such as distance 5 or 6.
This result suggests that the prefetch distance should not be 
chosen too small.

Another important parameter is the amount of data to be processed
for each ``prefetch-computation'' iteration.
In general, for an optimal use of the prefetch instruction,
the data stride per 
iteration should be longer than the length of a cache line.
Although it is easier to hide latency in the SSE macro 
with more data per iteration, too big data stride like 
the length of more than
four cache lines per iteration may reduce performance.
This is also dependent on the number of source fields in the 
application,
since the number of multiple prefetch requests is limited.
This number is  eight for the Opteron processor.

We perform several tests for each application and
determine the prefetch distance and data stride one by one.
For instance, in the routine to compute 
$\langle\psi,\phi\rangle$, data stride is the length of two cache
lines and prefetch both source fields at six cache lines ahead.
In the $\psi + c\phi$ routine, 
the data stride is the length of a cache 
line and the prefetch distance is five cache lines.

\begin{figure}[t]
    \includegraphics[width=7.9cm]{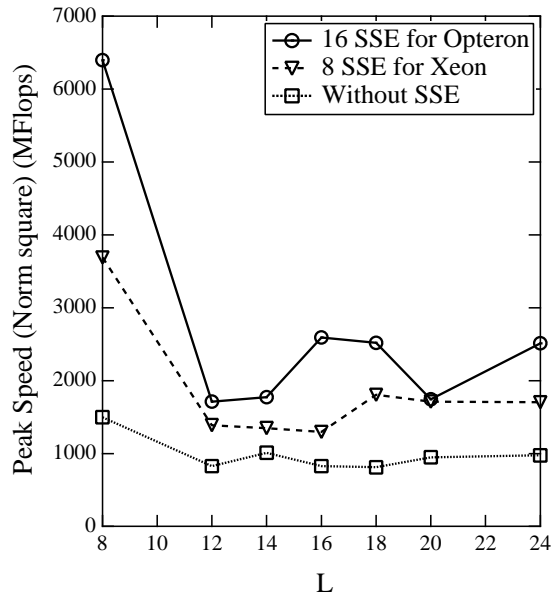}
    \caption{The lattice size dependence of the benchmark results
    for 32-bit $\langle\psi,\psi\rangle$ calculation written in C 
    language without SSE 
     ($\Box$), with SSE instructions optimized for Xeon ($\nabla$)
     and the new version optimized for Opteron ($\circ$). 
     All calculations are done on the Opteron cluster.
    }
    \label{fig:norm}
\end{figure}

\begin{figure}[t]
    \includegraphics[width=7.9cm]{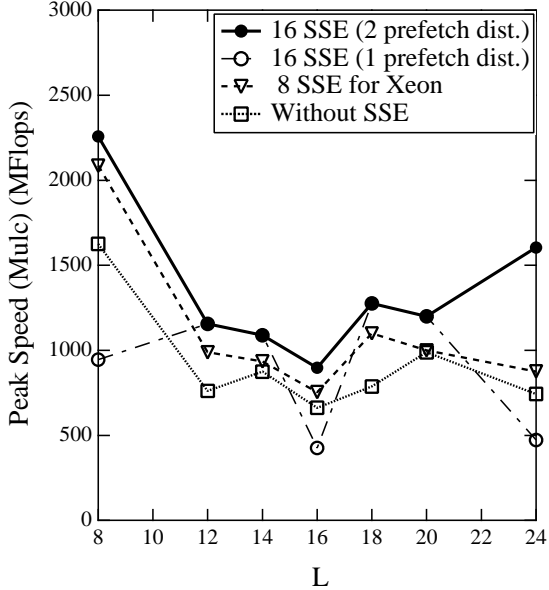}
    \caption{Benchmark results for 32-bit $ \psi+c\phi$. The slowing 
    down observed in the Opteron version ($\circ$) at $L=8,16,24$ 
    is cured of by adopting two prefetch distances ($\bullet$).}
    \label{fig:mulc}
\end{figure}

\section{Benchmarks}
\begin{figure}[t]
    \includegraphics[width=7.9cm]{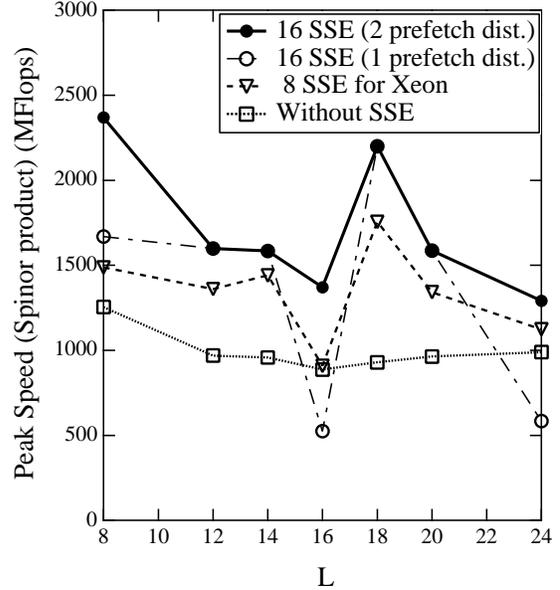}
    \vspace{-0.5cm}
    \caption{Benchmark results for 32-bit $ \langle \psi,\phi\rangle$. 
    The slowing 
    down observed in the Opteron version ($\circ$) at $L=16,24$ 
    is cured of by adopting two prefetch distances ($\bullet$).}
    \label{fig:spinor}
\end{figure}

To see the effectiveness of the implementation of SSE instructions 
and of optimization, we 
show benchmarks of the linear algebra routines with and without 
SSE instructions in Figs.~\ref{fig:norm},
\ref{fig:mulc} and \ref{fig:spinor}.
We also plot the result of the  original  code using 8 SSE resisters,
which is written for the Xeon processor 
and compatible with the Opteron processor.

The use of SSE instructions improves the performance even before
the optimization for the Opteron processor. 
Besides we have another considerable gain 
by tuning prefetch distance and by making 
use of 16 SSE registers.

The lattice size dependence of the performance is rather large for 
linear algebra routines.
Good performance  (6.4 GFlops) of 
$\langle\psi,\psi\rangle$ calculation at $L=8$ in Fig.~\ref{fig:norm} 
is due to the cache 
effect, where a spinor field 
(72 bytes $\times 8^{4}\sim 0.3 $ Mbytes) can 
entirely fit into the L2 cache.
We also observed slowing down in $ \psi+c\phi$ calculation
at $L=8, 16, 24$ as shown in Fig.~\ref{fig:mulc} and
in $ \langle \psi,\phi\rangle$ calculation at 
$L=16, 24$ as shown in Fig.~\ref{fig:spinor}.
They are
caused by bank conflict in multiple prefetch requests.
It happens for certain lattice volumes when two source fields are loaded 
as $\langle\psi,\phi\rangle$.
However,
this slowing down can be cured  by 
extending the prefetch distance of the second 
source by four, which is obtained by 
the maximal number of prefetch request divided 
by the number of source field, $8/2=4$. 
Such treatment is based on the observation  in Fig.~\ref{fig:prefetch}
that the longer prefetch 
distance does not affect performance.
The final version is roughly twice faster than the original C codes 
without SSE instructions.

In Table \ref{tbl:comp}, we compare the throughput of the Opteron 
processor to that 
of the Intel Xeon processor.
For the Dirac operator $Q\phi$, the ratio of the throughput is almost 
the same as that of the processor clock speed 
(2.4 GHz/1.7 GHz =1.41)
except for the 32-bit version at $L=8$, where the ratio is 1.64 $>$ 
1.41. 
This can again be the cache effect since a spinor 
and a gauge field can entirely
fit into the L2 cache ((72+96) bytes $\times 8^{4} 
\sim 0.7$ Mbytes).

\newcommand{\pmb}[1]{{#1}}
\begin{table}
    \caption{Benchmarks of AMD Opteron processor (2.4 GHz) and of Intel Xeon 
    processor (1.7 GHz) for the linear algebra routine and the Dirac 
    operator routine in units of Mflops. Ratio of performance is shown 
    in the last column, which can be compared to the processor clock
   speed ratio, 
    2.4/1.7=1.41.}
    \begin{tabular}{lccc}
	 \hline
	  & Opteron & Xeon & ratio  \\
	 \hline
	 $\pmb{ Q\phi}$ (32-bit, L=8) & 2462 & 1503 & 1.64 \\
	 $\pmb{ Q\phi}$ (32-bit, L=12) & 2037 & 1421 & 1.43 \\
	 $\pmb{ Q\phi}$ (64-bit, L=8) & 1140 & 799 & 1.43 \\
	 $\pmb{ Q\phi}$ (64-bit, L=12) & 1131 & 796 & 1.42  \\
	 $\pmb{\langle\psi,\phi\rangle}$ (32-bit, L=16) & 1370 & 617 & 2.22  \\
	 $\pmb{\langle\psi,\phi\rangle}$ (64-bit, L=12) & 1090 & 554 & 1.97  \\
	 $\pmb{\langle\psi,\phi\rangle}$ (64-bit, L=16) & 661 & 353 & 1.87  \\
	 \hline
     \end{tabular}
     \label{tbl:comp}
\end{table}

For linear algebra we have more gain as
expected from the difference of 
the processor clock speed. 
The larger data cache and the improvement of memory bandwidth 
may reflect the result
since linear algebra routines are rather memory limited.

\section{Summary}
In this report, we discuss the details of the optimization of the 
lattice QCD codes for the AMD Opteron processor.

Tuning of  prefetch distance for each application
can improve performance by more than  a 
factor of two.
The prefetch instruction, however, may cause significant slowing down 
in some linear algebra routines
at certain lattice sizes due to bank conflict. We found that
this slowing down can be cured  
by adopting two different prefetch distances for each source field.
The effect of the modification in SSE instructions can be observed 
only after the adjustment of the prefetch instructions.

Large cache effect is observed not only for the 
linear algebra routines but also for the
$Q\phi$ routines when the source fields can fit into 
relatively large L2 data cache  of the Opteron processor.
This suggests that 
the PC cluster of the Opteron processor can achieve
high performance when the sublattice size becomes small enough.
As a next step of this work, we will develop a parallel version of 
the codes.

\section{Acknowledgement}
The author would like to thank H. Wittig and Y. Koma for
a number of useful discussions. 
The author appreciate the original codes and the benchmark
programs written by 
L.~Giusti, M.~L\"uscher and H.~Wittig,  
which is the basis of the improvement
reported here.

The author would also like to thank
U. Ensslin of DESY IT group for technical supports.

\end{document}